# Ion-Assisted Nanoscale Material Engineering in Atomic Layers

Hossein Taghinejad,[1,2,3,] [*] Mohammad Taghinejad,[2, 4, 5] Sajjad Abdollahramezani,[2, 4] Qitong Li,[5] Eric V. Woods,[6, 7] Mengkun Tian,[6] Ali A. Eftekhar,[2] Yuanqi Lyu,[1] Xiang Zhang,[8] Pulickel M. Ajayan,[8] Wenshan Cai,[2] Mark L. Brongersma,[5] James G. Analytis,[1, 9, 10] [*] Ali Adibi[2, ][*]

1.  Department of Physics, University of California, Berkeley, CA, USA.
2.  School of Electrical and Computer Engineering, Georgia Institute of Technology, Atlanta, GA, USA.
3.  Kavli Energy NanoSciences Institute, University of California, Berkeley, CA, USA.
4.  School of Materials Science and Engineering, Stanford University, Stanford, CA, USA.
5.  Geballe Laboratory for Advanced Materials, Stanford University, Stanford, CA, USA.
6.  Institute of Electronics and Nanotechnology, Georgia Institute of Technology, Atlanta, GA, USA.
7.  Max Planck Institute for Iron Research, Max-Planck-Straße1, Düsseldorf, Germany.
8.  School of Materials Science and Nanoengineering, Rice University, Houston, TX, USA.
9.  Materials Sciences Division, Lawrence Berkeley National Laboratory, Berkeley, CA, USA.
10. CIFAR Quantum Materials, CIFAR, Toronto, Canada

*Corresponding Authors:
h.taghinejad@berkeley.edu
analytis@berkeley.edu
ali.adibi@gatech.edu

ABSTRACT. **Achieving deterministic control over the properties of low-dimensional materials with nanoscale precision is a long-sought goal. Mastering this capability has a transformative impact on the design of multifunctional electrical and optical devices. Here, we present an ion-assisted synthetic technique that enables precise control over the material composition and energy landscape of two-dimensional (2D) atomic crystals. Our method transforms binary transition metal dichalcogenides (TMDs), like $MoSe_2$, into ternary $MoS_{2\alpha}Se_{2(1-\alpha)}$ alloys with systematically adjustable compositions, $\alpha$. By piecewise assembly of the lateral, compositionally modulated $MoS_{2\alpha}Se_{2(1-\alpha)}$ segments within 2D atomic layers, we present a synthetic pathway towards the realization of multi-compositional designer materials. Our technique enables the fabrication of complex structures with arbitrary boundaries, dimensions as small as 30 nm, and fully customizable energy landscapes. Our optical characterizations further showcase the potential for implementing tailored optoelectronics in these engineered 2D crystals.**

The importance of synergy between geometric parameters and material composition for designing optoelectronic devices is exemplified in the development of heterostructure lasers. Double-heterostructure laser diodes, for example,[1,2] involve the heteroepitaxial encapsulation of a low-bandgap active material, often $Al_\alpha Ga_{1-\alpha}As$, between two large-bandgap materials, typically another $Al_\beta Ga_{1-\beta}As$ with a composition $\beta > \alpha$. The composition $\alpha$ governs the emission wavelength of the laser, and the geometry of the sandwiched configuration enables light confinement and efficient carrier recombination within the active region. Engineering the interplay between material composition and geometric facets paved the way towards creating quantum-well,[3,4] quantum-wire,[5] and quantum-dot lasers,[6] with superior performance, revolutionizing the field of optoelectronics.

Layered TMDs and their heterostructures are the newest class of optoelectronic materials in which their atomically thin nature yields novel quantum effects and unique device functionalities previously deemed impossible in bulk materials. Distinct properties such as strong excitonic effects,[7-9] fast carrier dynamics,[10,11] direct optical bandgap,[12] composition diversity ($MX_2$, where M: Mo, W and X: S, Se, Te),[13-15] and the possibility of co-integration on virtually any substrates give TMD heterostructures a competitive advantage in optoelectronic applications and quantum science. Despite the demonstrated potential, a material synthesis pathway towards programming the energy landscape and optoelectronic properties of TMD atomic crystals has remained an outstanding challenge, particularly at nanoscales. The widely adopted technique for preparing 2D heterostructures is the mechanical exfoliation and manual assembly,[16-18] which makes systematic control of geometric aspects and energy structure difficult. To gain better control over material properties, various methods have been explored for the direct growth of TMD heterostructures.[13,16,19-23] However, challenges such as random nucleation of heterojunctions,[14, 20] the restricting role of crystal symmetries on obtainable geometries,[24] and the lack of an engineerable tuning knob limit the capacity of direct-growth approaches in controlled formation of 2D materials with desired attributes. Further innovations such as modified edge-epitaxy,[25] laser templating,[26] and fast and automated control of reaction agents[27] have offered extra control. However, these developments target only specific material aspects, lacking a comprehensive solution for controlling the entire design parameters important in optoelectronics of 2D heterostructures. Thus, developing a holistic and scalable approach capable of simultaneously programming both geometric parameters and electronic structures of 2D heterostructures with nanoscale precision will be a key enabler for further advancements in 2D materials research.

Here, we introduce a synthetic pathway towards flexible engineering of in-plane properties in 2D TMDs, offering full control over geometric parameters (shape, dimensions, locations) and, independently, on material composition and bandgap energy without any inherent limitations imposed by geometry on composition or vice versa. Our technique is based on defect-mediated transformation of a binary TMD monolayer, such as $MoSe_2$, into a ternary $MoS_{2\alpha}Se_{2(1-\alpha)}$ alloy with a composition α that is tunable with the level of defects deliberately introduced into the host $MoSe_2$ film. We implement this strategy by introducing defects via focused-ion beam (FIB) irradiation of a monolayer $MoSe_2$ (Fig. 1a, middle) followed by low-temperature annealing in a sulfur-rich ambient (Fig. 1a, bottom). We show that the sulfurization step ensures exclusive introduction of sulfur into only ion-irradiated regions while preserving the composition of $MoSe_2$ films in pristine parts, enabling localized modulation of material properties within the 2D plane. The use of FIB offers full and independent control over geometry and energy landscape in two ways. Firstly, by merely modulating the ion dosage (D, $\frac{\#ions}{area}$), we exercise complete control over the material composition and bandgap energy of $MoS_{2\alpha}Se_{2(1-\alpha)}$ monolayers freely from α = 0 (i.e., $MoSe_2$, bandgap ≈ 1.5 eV) to α = 1 (i.e., $MoS_2$, bandgap ≈ 1.85 eV), enabling 350 meV bandgap modulation. Secondly, we leverage the facile beam scanning of the FIB to craft freeform heterostructures characterized by custom shapes and dimensions realized in desired locations. The dual capacity to control material composition and geometric parameters, independently and synergistically, enables the realization of designer optoelectronic, where complex energy landscapes can be synthetically dictated within atomically thin 2D materials.

Monolayer $MoS_{2\alpha}Se_{2(1-\alpha)}$ alloys can be synthesized via doping sulfur into pristine $MoSe_2$ films. This approach, however, requires annealing of $MoSe_2$ under sulfur gas at elevated temperatures (above ∼ 800 °C, Extended Data Fig. E1 and Fig. E2). However, we demonstrate that deliberate introduction of defects in a MoSe$_2$ host can lower the temperature requirement for sulfur doping and the synthesis of MoS$_{2\alpha}$Se$_{2(1-\alpha)}$ alloys. Accordingly, localized generation of defects via ion irradiation, followed by a low-temperature sulfurization, enables us to embed MoS$_{2\alpha}$Se$_{2(1-\alpha)}$ alloys within a host MoSe$_2$ monolayer and to create lateral heterostructures. The inset of Fig. 1b presents an optical image of a MoSe$_2$ monolayer that includes an ion-irradiated region (Methods section). As shown in Fig. 1b-I, irradiation induces photoluminescence (PL) quenching, implying non-radiative recombination of carriers via defect states. Generation of defects is further supported by Raman measurements (Fig. 1c-I), wherein the out-of-plane vibration of pristine MoSe$_2$ (A$_{1g, MoSe2}$) at ~241 cm$^{-1}$ undergoes broadening and a redshift to ~234 cm$^{-1}$. Additionally, two new spectrally broad features emerge around 145 cm$^{-1}$ and 293 cm$^{-1}$. Such spectral changes are attributed to the generation of defects and lattice disorder.[28] Defects result in softening of the restoring force acting on Mo–Se bonds, consequently causing a redshift in the A$_{1g, MoSe2}$ mode. Furthermore, disorder disrupts the selection rules, thereby giving rise to new Raman modes.

After sulfurization at 700 °C, the Raman mode of MoSe$_2$ disappears (Fig. 1c-II) and the in-plane (E$^2_{1g, MoS2}$, 383 cm$^{-1}$) and out-of-plane (A$_{1g, MoS2}$, 406 cm$^{-1}$) modes of MoS$_2$ appear, confirming the formation of Mo – S bonds within the irradiated region. However, the Raman spectrum of the pristine matrix surrounding the irradiated region still exclusively contains the A$_{1g, MoSe2}$ mode. This outcome proves the selective introduction of sulfur into ion-irradiated regions, which establishes a lateral MoSe$_2$ – MoS$_2$ heterostructure at the interface between pristine and ion-irradiated regions. However, the PL of the irradiated region still lacks any detectable signal after the sulfurization step (Fig. 1b-II), suggesting that the structure of the obtained MoS$_2$ is not fully ordered. To address this issue, we added one extra step after the sulfurization; we first purge the sulfur precursor and then subject the sample to in-situ annealing at a higher temperature (900 °C, 5 mins). In-situ annealing yields lattice reconstruction within the irradiated region, while the absence of sulfur gas prevents unintended doping in pristine regions. The 900°C used in this step is lower than the processing temperatures employed in alternative techniques.[26] Although follow-up studies are necessary to further reduce the synthesis temperature, the well-developed transfer techniques can facilitate the transfer of these engineered layers from the growth substrate to other substrates requiring low-temperature processes.

As illustrated in Fig. 1b-III, annealing restores PL emission within the irradiated region as manifested in the strong light emission at the $MoS_2$ bandgap of ~1.85 eV. Furthermore, optical reflectance measurements on the irradiated region reveal the appearance of characteristic A- and B-excitons of $MoS_2$ after the annealing step (Fig. 1d). $MoS_2$ Raman modes of the irradiated region also exhibit noticeable linewidth narrowing following the annealing step (Fig. 1c-III). Thus, Raman, PL, and optical reflection measurements confirm the effective recovery of the $MoS_2$ lattice following the in-situ annealing. We acknowledge that the annealing step sometimes induces slight changes in the PL and Raman spectrum of pristine regions, perhaps due to unwanted doping due to residual sulfur.

To gain a direct insight into structural transformations experienced by $MoSe_2$ films during the synthesis process, we conduct scanning transmission electron microscopy (STEM). We transfer $MoSe_2$ thin films onto a holey silicon-nitride grid (Fig. 2a) and irradiate an array of lines onto suspended sections (Fig. 2b). We then directly monitor the structural changes in the film via imaging the interface between ion-irradiated and pristine $MoSe_2$ before (Fig. 2c) and after (Fig. 2d) the sulfurization-annealing sequence. Fig. 2c indicates the amorphization of $MoSe_2$ following FIB irradiation, which explains the previously discussed suppression of PL emission and the redshift and broadening of Raman modes. *After* the sulfurization-annealing sequence (Fig. 2d), an ordered and multi-crystalline lattice is established within the irradiated region. This lattice reconstruction explains the recovery of PL emission, the appearance of excitonic signatures, and the narrowing of Raman linewidths, as we discussed above. We also note that in some scattered locations as exemplified in Fig. 2e, the lattice appears nearly completely reconstructed. However, the reconstruction in most locations reaches the level depicted in Fig. 2d. Collectively, we conclude that the localized modulation of the 2D material composition contains three steps (Fig. 2f); (i) introduction of Se-vacancy via ion irradiation, which also induces lattice deformation, (ii) infiltration of sulfur into Se-vacancies during the low-temperature sulfurization, and (iii) lattice reconstruction following the in-situ annealing step. We note that the observed amorphous-to-crystalline reconstruction has precedent in other chalcogenide-based compounds, most notably the family of germanium-antimony-telluride and its Se-doped variant (see also Extended Data).[29, 30]

A distinguishing advantage of our technique is the facile engineering of optoelectronic properties through tuning the material composition. To demonstrate this aspect, we perform the sulfurization-annealing sequence on a sample featuring ten rectangular regions exposed to ion dosages ranging from $D_1 = 0.1 \times 10^{13}$ cm$^{-2}$ to $D_{10} = 4.4 \times 10^{13}$ cm$^{-2}$. An optical image of a resulting sample is shown in Fig. 3a Raman measurements (Fig. 3b) demonstrate that at high ion dosages, Mo – S vibrations (i.e., $A_{1g, MoS2}$ and $E^2_{1g, MoS2}$) predominate the collected spectra, implying a composition close to binary MoS$_2$. In contrast, at low ion dosages, the Mo – Se vibration (i.e., $A_{1g, MoSe2}$) dominates, signaling that the lattice maintains a composition close to the original MoSe$_2$. At intermediate dosages, Raman spectra contain partial contributions from both Mo – S and Mo – Se bonds, indicating that a ternary MoS$_{2\alpha}$Se$_{2(1-\alpha)}$ alloy is established. Similarly, PL measurements, in Fig. 3c, demonstrate that as the ion dosage increases from $D_1$ to $D_{10}$, the PL peak energy continuously and monotonically blueshifts from the bandgap of MoSe$_2$ at 1.5 eV towards that of MoS$_2$ at ~1.85 eV. In essence, by merely tuning D we can precisely dictate the composition α and the bandgap energy (E$_g$) of embedded MoS$_{2\alpha}$Se$_{2(1-\alpha)}$ alloys within the host monolayer film (Fig. 3d). Our Raman characterizations (Fig. E3, Extended Data) attribute this systematic control to a varying level of ion irradiation-induced defects in host MoSe$_2$ monolayers.

The integration of FIB in our synthesis technique offers an advanced degree of control over energy landscape within the plane of 2D materials via freeform heterostructuring. We showcase this potential through realizing a complex heterostructure in the form of an ancient Lamassu, the Assyrian deity with the wings of a bird, the body of a lion, and the head of a human - in a literal sense, a "heterostructure" from antiquity. A schematic illustration of the spatial ion-dosage profile (details of FIB in Methods) and a representative optical micrograph of the printed pattern onto a monolayer MoSe$_2$ film are shown in Fig. 4a and Fig. 4b, respectively. For realizing the complex heterostructure, the sample subsequently undergoes the sulfurization-annealing sequence. Then, we used optically measured bandgaps to determine the composition α of MoS$_{2\alpha}$Se$_{2(1-\alpha)}$ compounds at different parts of the pattern and constructed a composition map in Fig. 4c. Three representative Raman spectra obtained from pure MoS$_2$, pure MoSe$_2$, and mixed Mo–S–Se regions, alongside their atomistic schematics, are presented in Fig. 4d. The correlation between the dosage map and the composition map demonstrates the successful synthesis of a complex multi-composition 2D heterostructure. The systematic composition variation in the wings and the large-scale uniformity in the body of the pattern corroborate the high degree of control that our technique offers. Such a piecewise assembly of arbitrarily shaped MoS$_{2\alpha}$Se$_{2(1-\alpha)}$ segments with programmable energy structures is an experimental confirmation of a synthetic pathway towards a designer material platform.

Here, we would like to highlight the advantages of our ion-assisted technique over alternative methods that generally involve patterned masking and atomic layer substitutions, as originally reported by Mahjouri et al.,[31] and further studied by us and others.[32, 33] As shown in Extended Data (Fig. E4), this approach requires multiple nanofabrication steps: (1) lithographic definition of patterns, (2) material deposition for hard masks, (3) chalcogen replacement, and (4) removal of patterned masks for electrical access to the heterojunction. In a multi-composition heterostructure, these four steps must be repeated for every $MoS_{2\alpha}Se_{2(1-\alpha)}$ segment with a different $\alpha$ value. Thus, such synthesis scheme results in the exponential growth of nanofabrication steps into an unfeasible number when targeting complex heterostructures. In sharp contrast, our approach enables advanced multi-composition heterostructuring (e.g., Fig. 4) in a single run without any complex gas delivery, material masking, or sequential nanofabrication steps. Additionally, removal of the hard mask for electrical access to the heterojunctions complicates the device fabrication flow in patterned masking approach. Our technique addresses this issue via eliminating hard masks form the synthesis process altogether. Finally, a key advantage of our approach is its fine spatial resolution. The ~30nm features demonstrated here are unattainable in the patterned masking technique due to the unwanted diffusion of reactive agents underneath the hard masks.

We next demonstrate that, despite the multi-crystalline nature, our engineered 2D materials yield junctions with optoelectronic properties suitable for photodetection and photovoltaic applications. We analyzed the alignment of electronic bands at a $MoS_2 - MoSe_2$ junction via photocurrent mapping experiments on a device shown in Fig. 5l. A representative photocurrent map under illumination with $\lambda = 550nm$ light and the drain-source voltage ($V_{ds}$) of 5V is shown in Fig. 5b. Comparing this photocurrent map with the optical reflection map, in Fig. 5a, reveals that the photocurrent is predominantly confined to the junction area. This implies that photogenerated electron-hole (e-h) pairs separate merely at the vicinity of the junction, indicating the type-II band alignment in which electrons and holes move to the opposite sides of the $MoS_2 - MoSe_2$ junction (Fig. 5k). As shown in Fig. 5b-e, photocurrent generation requires a threshold voltage of $V_{ds}^{thr} \approx$ 2V, as further highlighted in line-cuts across the $MoS_2 - MoSe_2$ interface (Fig. 5j). The possible role of the threshold voltage in breaking photo-generated excitons into free electron-hole pairs is discussed in Extended Data. As illustrated in Fig. 5b,g,i, increasing the wavelength of the excitation light from $\lambda = 550nm$ to 620nm and 730nm results in a monotonic reduction of the photocurrent across the junction. This trend stems primarily from a reduced band-to-band light absorption and inefficient e-h pair generation. The systematic dependence of photocurrent on photon energy further substantiates the creation of well-defined optoelectronic bandgaps in the constituting materials at the studied junction.

Our photocurrent mapping experiments collectively demonstrate that photocurrents are confined to the junction area, exhibit rectifying behavior, and reveal establishment of well-defined optical bandgaps in as synthesized heterostructures. These observations indicate that, despite interfacial disorders, the key characteristics of type-II band alignments in lateral junctions are preserved in heterostructures produced by our ion-assisted approach. Thus, while the full crystallinity of lateral junctions is preferred, it is not a requirement for impacting the photovoltaic applications of 2D materials. This conclusion is technologically significant. For instance, in the mainstream silicon (Si) photovoltaic technology the optimal efficiency of single crystalline Si is compromised for the benefit of lower-cost production with *polycrystalline* Si. Similarly, the ion-assisted technique yields optoelectronic-grade 2D heterostructures in which the lack of full junction-crystallinity is balanced against the extra advantages of finely tunable bandgap energy as well as designable geometric aspects of 2D junctions.

In conclusion, our introduced ion-assisted synthetic approach offers flexible control over properties of 2D layers beyond their intrinsic forms. The ability to locally modulate material composition within complex geometric profiles grants material designers full access to the entire parameter space, enabling the conception of more advanced device structures. On one hand, the synthesis technique is scalable to large areas, only limited by the size of the host 2D materials (approximately 100 μm in our samples, Fig. 4). On the other hand, the spatial resolution of FIB microscopes allows for nanometer-scale precision, as demonstrated by our ability to achieve dimensions as small as tens of nanometers in suspended films (Fig. 2b). Replacing $Ga^+$ with $He^+$ ions can further push boundaries to dimensions approaching 0.5nm.[34] Capitalizing on this resolution, our ion-assisted method offers a potentially transformative pathway towards unlocking novel quantum-confinement effects that emerge at length scales approaching the exciton Bohr radius of 2D TMDs (a few nanometers). While ion-substrate interactions enlarge the irradiated area,[35] the bending of electronic bands at the lateral heterojunctions can tighten spatial confinement to dimensions smaller than metallurgical dimensions. This further underscores another facet of our approach, that is the full control over tuning of electronic structure via merely adjusting the ion dosage. Such distinct features offer a promising pathway towards the scalable realization of quantum dots and quantum wells in deterministic and adaptable fashions with tunable optoelectronic properties via facile design of geometry and energy landscape.

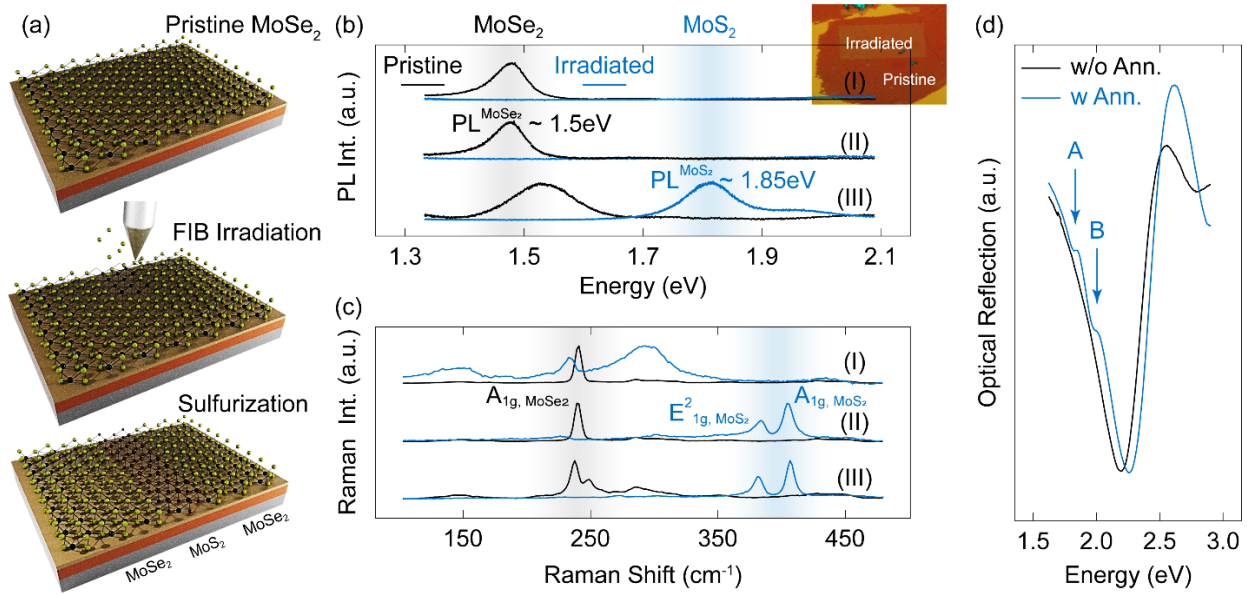

**Figure 1: Ion-Assisted Composition Modulation of 2D Layers.** (a) Schematic representation of the synthesis protocol. (b, c) Normalized PL and Raman spectra of pristine and irradiated (D ≈ 2.5 × 10¹³ cm⁻²) MoSe₂ at three stages: (I) before sulfurization, (II) after sulfurization, and (III) after sulfurization-annealing sequence. Sulfurization and annealing temperatures are 700 ºC (10 mins) and 900 ºC (5 mins), respectively. Inset in panel (b) shows the optical image of the FIB-irradiated sample. (d) Optical reflectance obtained from two sulfurized samples with and without in-situ annealing.

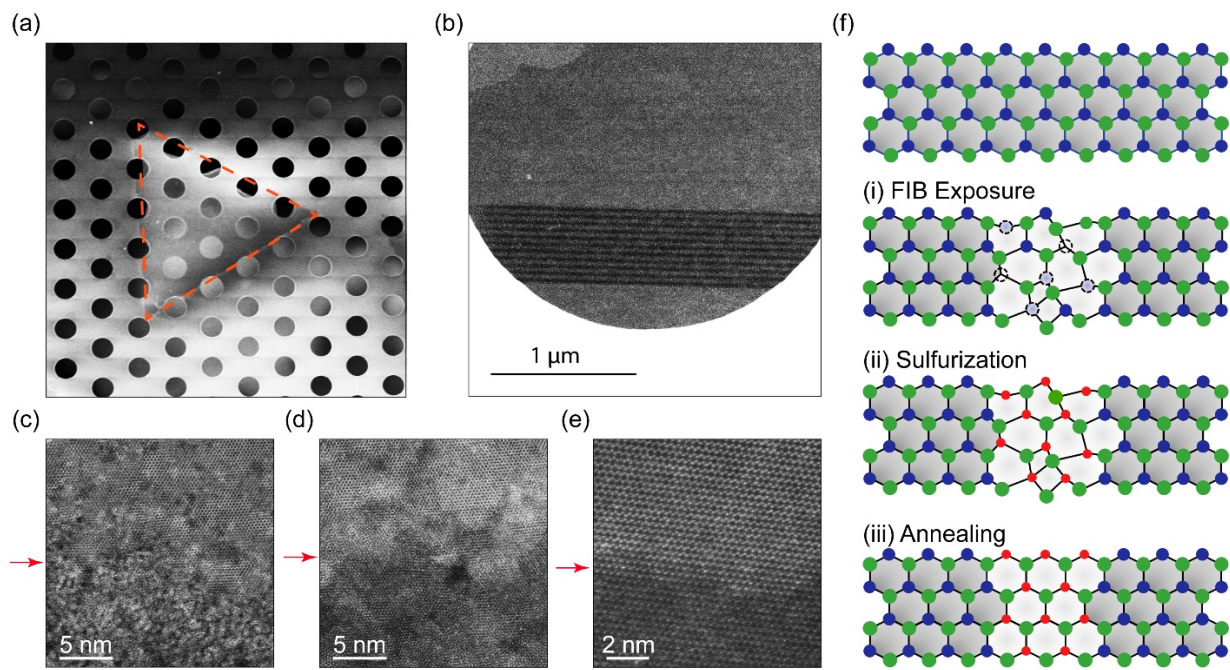

**Figure 2: Controlled Restoration of Lattice Structure.** (a) Low-magnification STEM image of a few-layer $MoSe_2$ film transferred onto a holey silicon nitride grid. The $MoSe_2$ crystal is outlined with a dashed triangle. (b) STEM image of an array of lines (width $\approx$ 30nm, pitch $\approx$ 50nm) FIB patterned onto a part of the $MoSe_2$ crystal that is suspended over a hole. (c, d) High-resolution STEM images at interfaces between FIB-irradiated and pristine $MoSe_2$ obtained before and after the sulfurization-annealing sequence, respectively. Arrows point to the interface, and the bottom half of the image is irradiated. (e) A sample interface showing occasional near-complete reconstruction of the lattice. (f) Schematic illustration of structural evolutions at different stages; (i) formation of vacancies (dotted circles) and crystal deformation after FIB irradiation, (ii) selective incorporation of sulfur into the irradiated region after low-temperature sulfurization, and (iii) reconstruction of the lattice after the high-temperature annealing. Green, blue, and red circles represent Mo, Se, and S atoms, respectively.

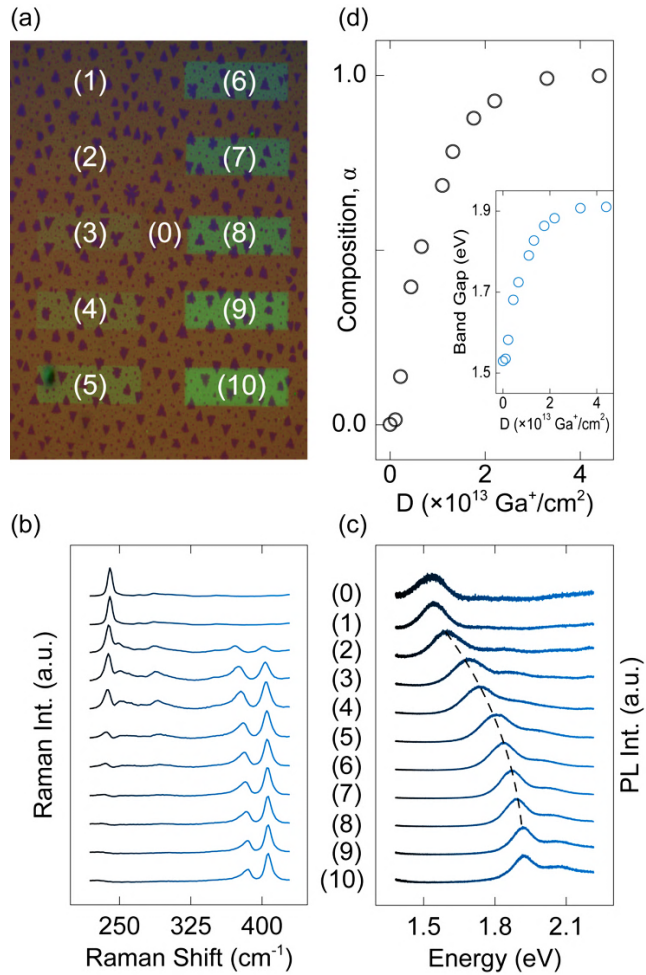

**Figure 3: Continuous Control of Bandgap Energy.** (a) Optical image of a representative sample after the sulfurization-annealing sequence. Rectangles indexed as (1) – (10) are irradiated with ion dosages $D_1 - D_{10} = 0.1 - 4.4$ ($\times 10^{13}$ cm$^{-2}$). Each rectangle is $8 \times 22$ μm$^2$. The region marked as (0) is pristine MoSe$_2$. (b, c) Normalized Raman and PL spectra obtained from regions indexed on panel (a). For clarity, spectra are displaced vertically. The dashed line on (c) is a guide to eyes, connecting PL peak positions. (d) Composition $\alpha$ and bandgap energy (inset) of MoS$_{2\alpha}$Se$_{2(1-\alpha)}$ compounds as a function of the ion dosage.

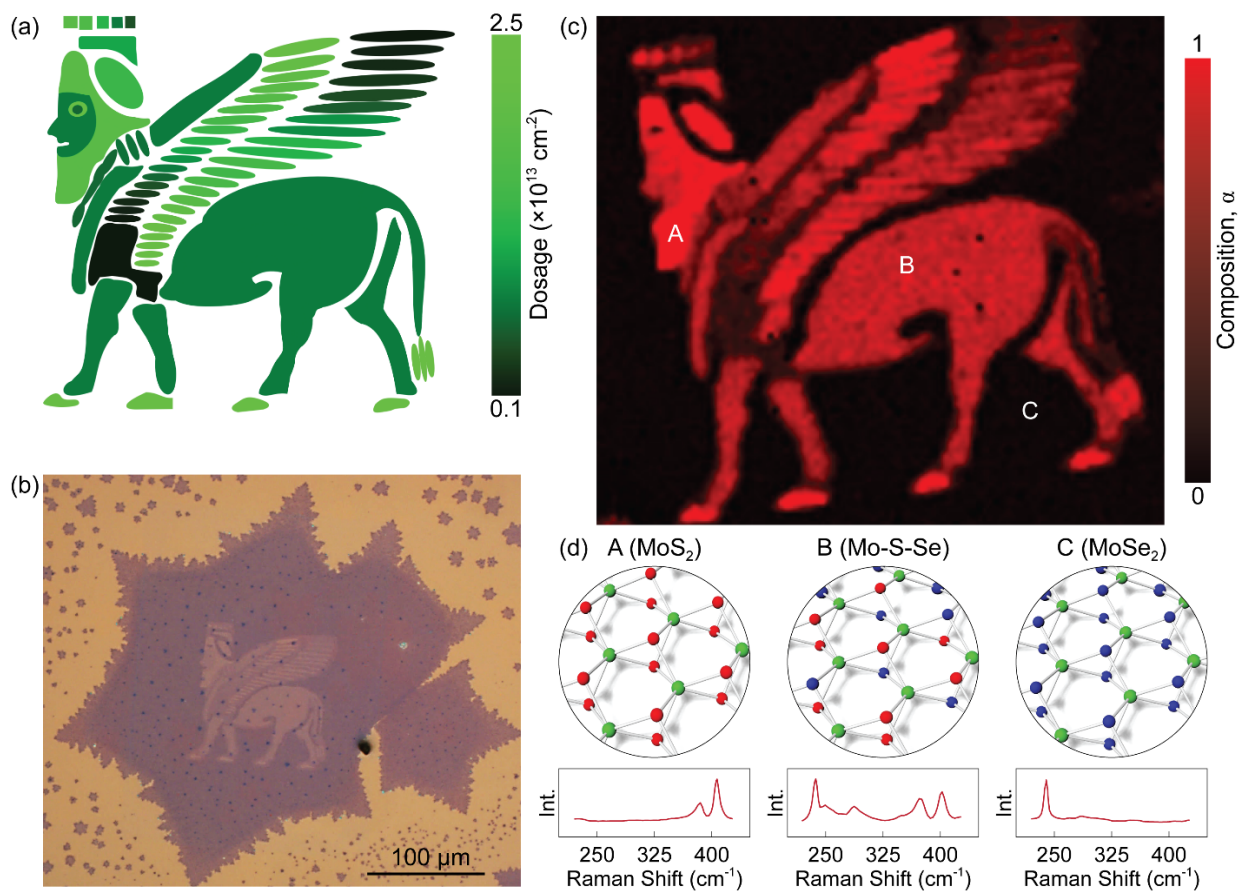

**Figure 4: Multi-Composition Designer Materials.** (a) Schematic drawing of ancient Lamassu to be FIB-printed onto a $MoSe_2$ monolayer. Different ion dosages are represented by different shades of green. (b) Optical micrograph of a representative sample that shows the pattern is FIB-printed onto the $MoSe_2$ monolayer. (c) Two-dimensional composition map of the complex in-plane heterostructure established after the sulfurization-annealing sequence. (d) Atomistic schematics (top) and Raman spectra (bottom) of the three points marked on the composition map. Established compounds at points A, B, and C are $MoS_2$, $Mo-S-Se$, and $MoSe_2$, respectively. Green, red, and blue spheres represent Mo, S, and Se, respectively.

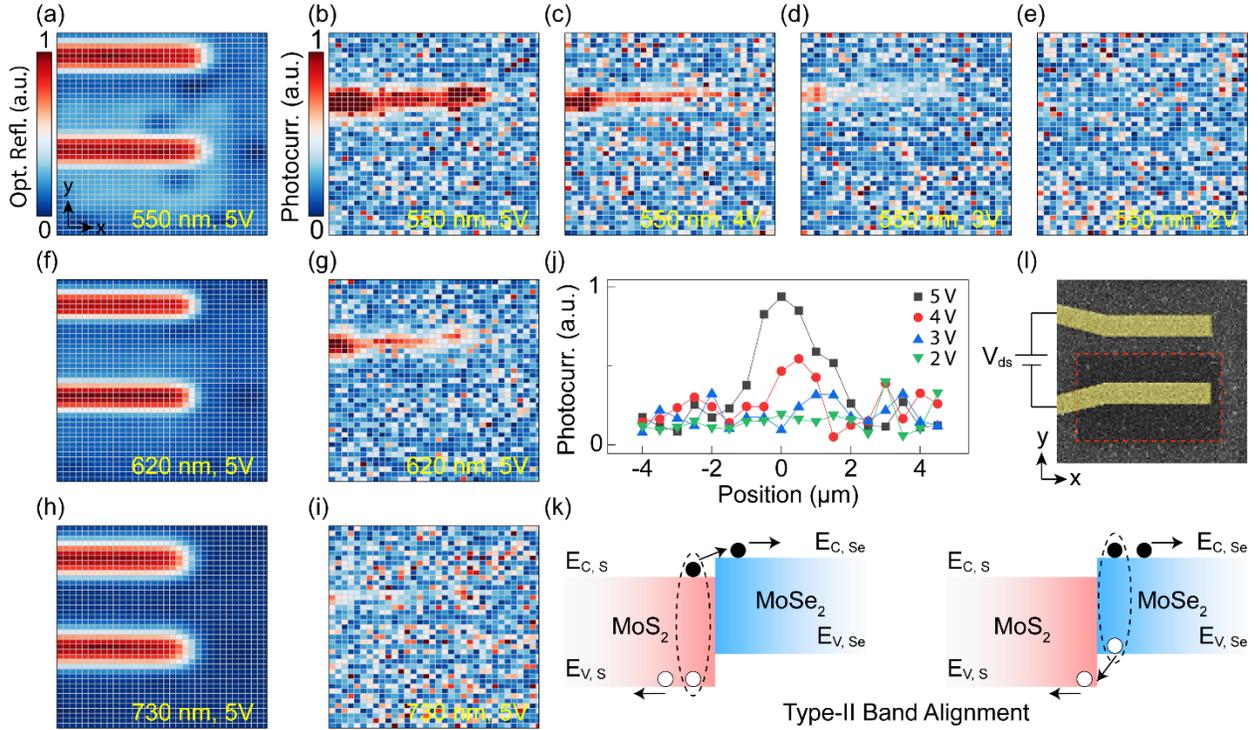

**Figure 5: Demonstration of Photovoltaic Effects.** (a) Optical reflection map obtained under illumination with 550nm light. Drain-source electrodes (Au/Ti) are identified by strong optical reflections. (b-e) Photocurrent maps obtained under fixed wavelength $\lambda = 550$nm and $V_{ds} = 5$, 4, 3, and 2V, respectively. Optical reflection and photocurrent maps obtained under (f, g) ($\lambda = 620$nm, $V_{ds} = 5$V) and (h, i) ($\lambda = 730$nm, $V_{ds} = 5$V). (j) Line cuts of photocurrents across the heterojunction interface at different $V_{ds}$ values and a fixed $\lambda = 550$nm. The junction is located at 0µm. (k) Schematic representation of e-h pair separation across a type-II lateral heterojunction. $E_c$ and $E_v$ represent the conduction- and valance-band edges, respectively. Open and closed circles represent holes and electrons, respectively. (l) A representative SEM image of the measured device. The dashed box outlines the region converted into $MoS_2$. Maps are obtained from an area of 17µm × 20µm. Photocurrent maps are normalized to the strongest current measured among all ($\lambda$, $V_{ds}$) combinations. Optical reflection maps are normalized separately for each $\lambda$.

## Methods

**Focused-ion-beam irradiation.** Ion irradiation experiments were conducted in a dual-column Gallium ion (Ga$^+$) FIB/SEM microscope (FEI Nova NanoLab 200) at a fixed acceleration voltage of 5kV and varying ion-beam currents (I) of 5pA, 20pA, or 80pA. To ensure uniform exposure in large-area irradiations, we intentionally increase the beam area (A) by introducing defocusing. We also deliver an intended ion dosage in multiple passes (N$_p$). Accordingly, the ion dosage is calculated as D = $\frac{It_dN_p}{qA}$, where $t_d$ and $q$ are the dwell time and the Ga$^+$ charge, respectively.[36] Sample navigations and imaging were conducted using the electron beam (i.e., the SEM mode). For experiments presented in Fig. 4, we rely on the pattern generator of the FIB system to scan the ion beam over an arbitrarily complex spatial profile defined within an input bitmap file. Guided by the bitmap file, the pattern generator sends signals to the electrostatic deflection plates inside the FIB column for vector-scanning the ion beam over MoSe$_2$ monolayers. Each pixel in the bitmap file contains a combination of red, green, and blue (R, G, B) integer values, each varying between 0 and 255. The R content is not currently used by our pattern generator, and we keep it constant at 0. The G content determines the dwell time $t_d$, which in turn defines the delivered ion dosage (for a given $I$ and $N_p$) and eventually the composition of the MoS$_{2\alpha}$Se$_{2(1-\alpha)}$ lattice at each pixel. $t_d$ is 0.1 μSec at G = 0, and it linearly increases with increasing the G value. Finally, the B content controls the beam blanker, and for any nonzero value the beam is unblanked. Accordingly, we use the (0, G, 255) combination for pixels residing within our desired pattern and (0, 0, 0) everywhere else.

**Sulfurization and annealing processes.** For the sulfurization step, we load sulfur powder into a quartz crucible and place it next to SiO$_2$/Si substrates covered with CVD-grown MoSe$_2$ monolayers. Then we warm up the chamber from room temperature to a targeted sulfurization temperature and hold it there for approximately 10mins. For annealing steps, we first flush the reaction chamber by steadily flowing Ar gas into the chamber to deplete sulfur gas. Then, temperature is elevated to ~900 °C for annealing the sample for ~5mins. Eventually, samples are cooled to room temperature.

**Optical characterization.** PL and Raman measurements were performed using a Renishaw inVia confocal microscope with a laser excitation wavelength of 532 nm focused down to a ~1 μm diameter spot using a 100X objective lens. The laser power is kept below 100 μW to mitigate laser heating. The composition α of $MoS_{2\alpha}Se_{2(1-\alpha)}$ compounds (shown in Fig. 3d) is extracted from optical bandgaps ($E_g$) in PL measurements. Assuming a linear modulation with negligible band-bowing,[37] we extract α from $E_{g, \alpha} = (\alpha) E_{g, MoS2} + (1-\alpha) E_{g, MoSe2}$. Optical reflection measurements are performed at a normal incident angle using a 50X objective lens with a numerical aperture of ~ 0.5. A tungsten halogen lamp is used as the source, and a Craic QDI 202 micro-spectrophotometer mounted on a Leica DM 4000M microscope serves as the detector.

**Photocurrent mapping**. Photocurrent mappings are carried out in a custom-built optoelectronic setup. A supercontinuum laser with two acousto-optic tunable filters (Fianium) is used to tune the wavelength of monochromatic light used for photoexcitation of 2D heterostructures. A 5X long-working-distance objective lens (Mitutoyo M Plan APO, 0.14 NA) focuses the light onto the sample. To image the sample, two 50:50 beam splitters, a halogen lamp with a diffuser, and a CCD imaging camera with tube lenses are placed before the objective. Samples are wire-bonded to a chip carrier and then mounted on a three-axis piezo stage with a rotating platform to accurately focus the beam spot at the center of the devices. A source meter (Keithley 2612) is used to extract the photocurrent from the sample biased at different $V_{ds}$ voltages.

**STEM imaging.** The STEM imaging is performed in an 80 kV aberration-corrected Hitachi HD2700 microscope operated in the high-angle annular dark field (HAADF) mode. Beam current is kept below 30 pA to minimize the ionization damage. To enable high temperature processes, MoSe$_2$ films are transferred onto holey grids made of SiN.

**Device fabrication.** Standard electron-beam lithography (EBL) using Poly(methyl methacrylate) (PMMA) is used for definition of electrical contacts. Then, Au/Ti (60nm/10nm) metal contacts are deposited via e-beam evaporation, followed by a liftoff in acetone for resolving contacts. To ensure Ohmic contacts, devices are annealed at 200 $^o$C under the flow of H$_2$ gas diluted in Ar.

## Data availability

The data that support findings of this study are available from the corresponding authors on reasonable request.

## Acknowledgment

This work is primarily funded by the Air Force Office of Scientific Research (AFOSR) under Grant No. FA9550-15-1-0342 (G. Pomrenke). HT acknowledges the financial support of the Kavli Energy NanoScience Institute (ENSI) through the Heising-Simons Postdoctoral Fellowship at the University of California, Berkeley. JGA acknowledges support of the Heising-Simons Faculty award H5881 and LBNL Quantum Systems Accelerator, funded by the U.S. Department of Energy, Office of Science, National Quantum Information Science Research Centers. MT and MLB acknowledge support from the U.S. Department of Energy (DE-FG07-ER46426). HT and AA acknowledge technical support from Georgia Tech Institute for Electronics and Nanotechnology (IEN). IEN is a member of the National Nanotechnology Coordinated Infrastructure and supported by the National Science Foundation under Grant No. ECCS-1542174.